\documentclass[aps,prl,superscriptaddress, twocolumn]{revtex4-1}
\usepackage{graphicx}
\usepackage{dcolumn}
\usepackage{bm}
\usepackage{amsmath}

\usepackage{color}
\usepackage{bm,graphicx,hyperref}
\hypersetup{%
  breaklinks = {true},
  citecolor = {blue},
  colorlinks = {true},
  linkcolor = {red},
}

\begin{document}

\title{Localized Magnetic States in 2D Semiconductors}

\author{A. S. Rodin} \affiliation{Centre for Advanced 2D
  Materials, National University of
  Singapore, 6 Science Drive 2, 117546, Singapore}

\author{A. H. Castro Neto}
\affiliation{Centre for Advanced 2D
  Materials, National University of
  Singapore, 6 Science Drive 2, 117546, Singapore}

\date{\today}

\begin{abstract}

We study the formation of magnetic states in localized impurities embedded into two-dimensional  semiconductors. By considering various energy configurations, we illustrate the interplay of the gap and the bands in the system magnetization. Finally, we consider finite-temperature effects to show how increasing $T$ can lead to formation and destruction of magnetization.

\end{abstract}

\maketitle

Over fifty years ago, Philip W. Anderson proposed a model bearing his name to describe magnetic impurities in metals.~\cite{Anderson1961lms} In his seminal paper, he used the mean-field approach to show how an embedded impurity can become magnetized if certain requirements are met. Since then, this model, whose wide applicability is matched by its  elegance, has been used extensively in the fields of heavy fermions~\cite{Fulde1988itt} and the Kondo effect~\cite{Kondo1964rmi}. Several years ago, it was used to describe the magnetization of localized impurities in graphene.~\cite{Uchoa2008lms} In this work, we extend the analysis to include two-dimensional semiconductors with Dirac-like dispersion.

Advances in fabrication and manipulation of low-dimensional materials have yielded a number of experiments studying the magnetic nature of 2D systems. For example, it has been demonstrated that vacancies and atomic adsorbates can give rise to magnetized states in graphene.~\cite{ Ugeda2010maa, Nair2012shp, McCreary2012mmf, GonzalezHerrero2016asc} It has also been shown that defects in graphene can lead to the Kondo effect~\cite{Chen2011tke}, a clear signature of the magnetic states. In addition, novel two-dimensional magnetic materials have recently been reported~\cite{Gong2017doi, Huang2017ldf}, giving rise to an entirely new direction on condensed matter physics. Given the growing interest in the field, our goal is to provide a clear and intuitive understanding of magnetic effects arising in localized states in the presence of a gap.

Here, we use the Anderson's model to explore the rich magnetization phase space of a 2D semiconductor. In contrast to earlier studies which focused on the zero-$T$ regime, we explicitly include temperature in our work. By performing the calculations at finite $T$, we demonstrate that magnetization has a non-trivial dependence on temperature. Even though our work focuses on the massive Dirac systems, the qualitative results are general and applicable to a wide range of the ever-growing members of the 2D material family.

We begin by constructing a Hamiltonian following the method laid out by Anderson. The first component describes the bulk system:
\begin{equation}
	H_E = \sum_{jq\sigma} \left(E_{jq\sigma} - \mu\right)c_{jq\sigma}^\dagger c_{jq\sigma}\,,
	\label{eqn:H_E}
\end{equation}
where the indices $j$, $q$, and $\sigma$ label the band, momentum, and spin, respectively, and $\mu$ is the chemical potential. At this point, we make no assumptions about the dimensionality of the system or the distribution of the energy levels.

The second part of the Hamiltonian introduces a localized impurity state at the origin:
\begin{equation}
	H_I = \mathcal{E} \sum_\sigma a^\dagger_\sigma a_\sigma + U a_\downarrow^\dagger a_\uparrow^\dagger a_\uparrow a_\downarrow\,.
\label{eqn:H_I}	
\end{equation}
Here, $\mathcal{E}$ is the spin-independent on-site energy and $U$ is the Coulomb term arising from the electron-electron repulsion. In accordance with the Anderson's prescription, we use the mean field approximation to rewrite the quartic term in Eq.~\eqref{eqn:H_I} as $a^\dagger_\uparrow a^\dagger_\downarrow a_\downarrow a_\uparrow \approx \langle n_\uparrow\rangle a^\dagger_\downarrow a_\downarrow  + a^\dagger_\uparrow a_\uparrow \langle n_\downarrow \rangle - \langle n_\uparrow
 		\rangle \langle n_\downarrow \rangle$.

Finally, we couple the bulk and the impurity using a hybridization term:
\begin{align}
	H_C &= V\sum_{j\sigma} \left[a_\sigma^\dagger c_{j\sigma}\left(0\right) + c^\dagger_{j\sigma}\left(0\right) a_\sigma\right]=
	\nonumber
	\\
	&= \frac{V}{\sqrt{N}}\sum_{jq\sigma} \left[a_\sigma^\dagger c_{jq\sigma} + c^\dagger_{jq\sigma} a_\sigma\right]\,,
	\label{eqn:H_C}
\end{align}
where $N$ is the number of states in the Brillouin zone.

Combining the Hamiltonians from Eqs.~\eqref{eqn:H_E}-\eqref{eqn:H_C} yields the imaginary-time action
\begin{align}
	S &= -\beta U\langle n_\uparrow
 		\rangle \langle n_\downarrow \rangle + \sum_{n\sigma}\bar\phi_{n\sigma}\left(-i\omega_n+\mathcal{E}_\sigma\right)\phi_{n\sigma}
 		\nonumber
 		\\
 		&+ \sum_{njq\sigma}\bar\chi_{njq\sigma}\left(-i\omega_n+E_{jq\sigma}-\mu\right)\chi_{njq\sigma} 
 		\nonumber
 		\\
 		&+ \frac{V}{\sqrt{N}}\sum_{njq\sigma}\left(\bar\phi_{n\sigma}\chi_{njq\sigma} + \bar\chi_{njq\sigma}\phi_{n\sigma}\right)\,.
\label{eqn:S}
\end{align}
Here, $\chi$ and $\phi$ are the Grassmann fields corresponding to the operators from the Hamiltonians. In addition, we have defined the total energy for an electron of spin $\sigma$ at the impurity as $\mathcal{E}_\sigma = \mathcal{E} + U\langle n_{-\sigma}\rangle$.

From the action, we obtain the partition function:
\begin{align}
	\mathcal{Z} & = \int \mathcal{D}(\bar\chi,\chi)\, \mathcal{D}(\bar\phi,\phi)e^{-S} 
	\nonumber
	\\
	&=e^{\beta U\langle n_\uparrow
 		\rangle \langle n_\downarrow \rangle}\mathcal{Z}_E \prod_{n\sigma}\left[-\beta \mathcal{G}_\sigma^{-1}\left(i\omega_n\right)\right]\,,
 	\label{eqn:Z}
 	\\
 	\mathcal{G}_\sigma^{-1}\left(z\right)&=z-\mathcal{E}_\sigma+\frac{V^2}{N}\sum_{jq}\frac{1}{-z+E_{jq\sigma}-\mu}\,.
 	\label{eqn:Inv_G_z}
\end{align}
The quantity $\mathcal{Z}_E$ is the partition function for the independent electron system and $\mathcal{G}_\sigma(z)$ is the Green's function for spin $\sigma$ at the impurity. Note that the nature of the states of $H_E$ is not important; only the energies and the degeneracies of these levels matter. The occupation number for the spin state can be obtained from
\begin{equation}
	\langle n_\sigma\rangle = \frac{1}{\beta}\sum_n \mathcal{G}_\sigma\left(i\omega_n\right)\,,
	\label{eqn:n_sigma}
\end{equation}
where $\omega_n$ are the fermionic Matsubara frequencies. Equation~\eqref{eqn:n_sigma} shows that $\langle n_\sigma\rangle$ is a function of $\mathcal{E}_\sigma = \mathcal{E} + U\langle n_{-\sigma}\rangle$. Therefore, to calculate the occupation numbers for the spins one must solve a pair of non-linear equations. For a non-magnetized system, there is a single solution with $\langle n_{\uparrow}\rangle = \langle n_\downarrow\rangle$. In the magnetized case, there are three solutions: one with equal occupation and the other two with $\langle n_{\uparrow}\rangle \neq \langle n_\downarrow\rangle$ which are lower in energy.~\cite{Anderson1961lms}


Having set up the formalism, we turn to a concrete example of a massive Dirac dispersion given by 
$\pm\sqrt{m^2 + \gamma^2 q^2}$. This results in
\begin{equation}
	\mathcal{G}_\sigma^{-1}\left(z\right) = z - \mathcal{E}_\sigma +\left(z + \mu\right) \Delta \ln\left[\frac{D^2}{m^2 - \left(z+ \mu\right)^2}+1\right]\,,
	\label{eqn:Inv_G_Dirac}
\end{equation}
where $\Delta = V^2/(2\pi D)$ and $D$ defines the energy cutoff as $\pm\sqrt{m^2 + D^2}$. To make the frequency summation in Eq.~\eqref{eqn:n_sigma} more convenient, we define
\begin{equation}
	\mathcal{P}_\sigma^{-1}\left(z\right) = z - \frac{\epsilon_\sigma}{D} + z\Delta\ln\left(\frac{1}{\frac{m^2}{D^2}-z^2}+1\right)\,,
	\label{eqn:P}
\end{equation}
with $\epsilon_\sigma = \mathcal{E}_\sigma+ \mu$ (and, similarly, $\epsilon = \mathcal{E} + \mu$). This allows us to rewrite the occupation number as
\begin{equation}
	\langle n_\sigma\rangle = \frac{1}{D\beta}\sum_n\mathcal{P}_\sigma\left(\frac{i\omega_n+\mu}{D}\right)\,.
	\label{eqn:n_from_P}
\end{equation}

The function $\mathcal{P}_\sigma\left(z\right)$ has poles and branch cuts on the real axis. The cuts are found in the regions where the argument of the logarithm is negative: $z \in \left(m/D,\sqrt{1 + m^2/D^2}\right)$, $\left(-\sqrt{1+m^2/D^2},-m/D\right)$. The poles are located at the zeros of $\mathcal{P}_\sigma^{-1}\left(z\right)$ of which there are three: one below both branch cuts, one above, and one between them.

In accordance with the Matsubara method, we need a function with poles at $\left(i\omega_n + \mu\right)/D$ with residue $-1/\left(D\beta\right)$. Recalling that $\omega_n = \left(2n + 1\right)\pi/\beta$, we get $f\left(z\right) = \left[e^{D\beta\left(z-\frac{\mu}{D}\right)}+1\right]^{-1}$, which is simply the Fermi-Dirac distribution with the energies divided by $D$.

Following the standard procedure of contour integration with branch cuts and poles leads to
\begin{widetext}
\begin{align}
	\langle n_\sigma\rangle & =\sum_j W_\sigma\left(\zeta_j\right)f\left(\zeta_j\right) - \int_{-\sqrt{1+\frac{m^2}{D^2}}}^{\sqrt{1+\frac{m^2}{D^2}}}\frac{f\left(z\right)\text{Im}\left[\mathcal{P}_\sigma\left(z+i0\right)\right]\Theta\left(\left|z\right|-\frac{m}{D}\right)}{\pi}dz\,,
	\\
	W_\sigma^{-1}\left(z\right) &= \frac{d}{dz}\mathcal{P}_\sigma^{-1}\left(z\right) = 1 + \Delta\left[\ln\left(\frac{1}{\frac{m^2}{D^2}-z^2}+1\right) + \frac{2z^2}{\left(\frac{m^2}{D^2}-z^2+1\right)\left(\frac{m^2}{D^2}-z^2\right)}\right]\,,
	\\
	\text{Im}\left[\mathcal{P}_\sigma\left(z+i0\right)\right] &= \frac{-\pi\Delta\left|z\right|}{\left[z-\frac{\epsilon_\sigma}{D}+z\Delta\ln\left(\frac{1}{z^2-\frac{m^2}{D^2}}-1\right)\right]^2+\pi^2\Delta^2z^2}\,,
\end{align}	
\end{widetext}
where the index $j$ runs over the poles of $\mathcal{P}_\sigma\left(z\right)$.

At this point, we can proceed to the main task at hand: determining the magnetization of the system. We approach the problem systematically by addressing different energy configurations separately. For now, we set the temperature to zero.

The first configuration we consider is the one where the empty impurity state is in the gap and $\epsilon + U < m$. An equivalent way of stating this is $|\epsilon_\sigma| < m$. This means that the gap pole carries a substantial amount of the spectral weight since it is close to the original level in energy. It is possible to show from Eq.~\eqref{eqn:P} that the gap pole will always be shifted towards the middle of the gap with respect to $\epsilon_\sigma$. The reason lies in the interaction with the symmetric conduction and valence bands: the level is repelled more strongly by the band it is closer to.

Despite the pole shift, one might be tempted to draw an analogy with an isolated impurity and conclude that since $\epsilon_\sigma$ is in the gap, the impurity is magnetized whenever the chemical potential is between the shifted values of $\epsilon$ and $\epsilon + U$. This, however, results in an overestimation of the magnetic range of $\mu$. To understand why this is it the case, let us temporarily ignore the energy shift and rewrite the singly- and doubly-occupied single-particle energies as $\epsilon_\sigma = \epsilon + U n_\text{min}$ and $\epsilon + U n_\text{max}$. For an isolated impurity, the minimum and maximum occupation numbers are 0 and 1, respectively. Here, however, because of the spectral tails in the bands, as long as $\mu$ is inside the gap, the occupation number range is smaller than $[0,1]$.

\begin{figure}[h]
	\includegraphics[width = 1.65in]{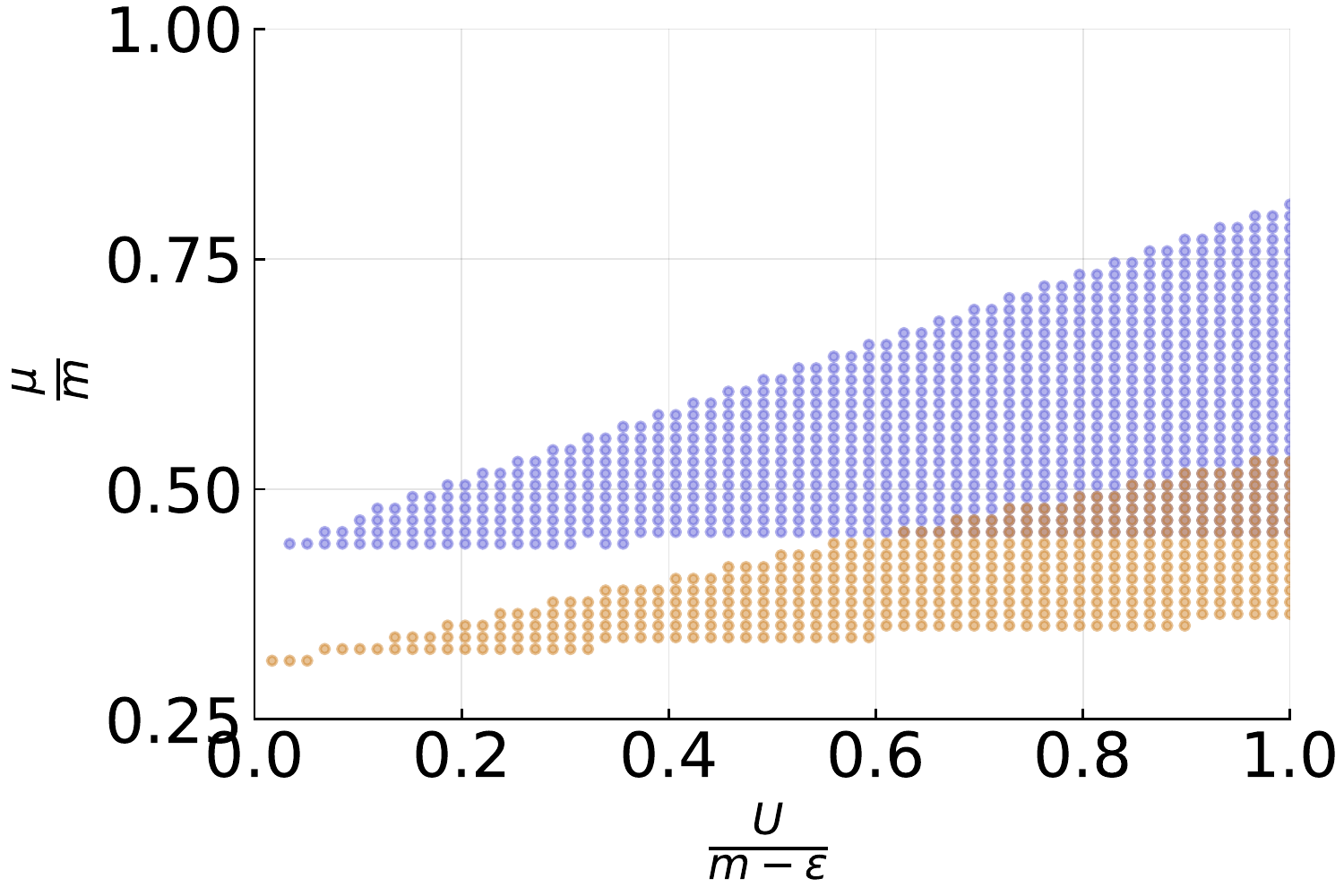}
	\includegraphics[width = 1.65in]{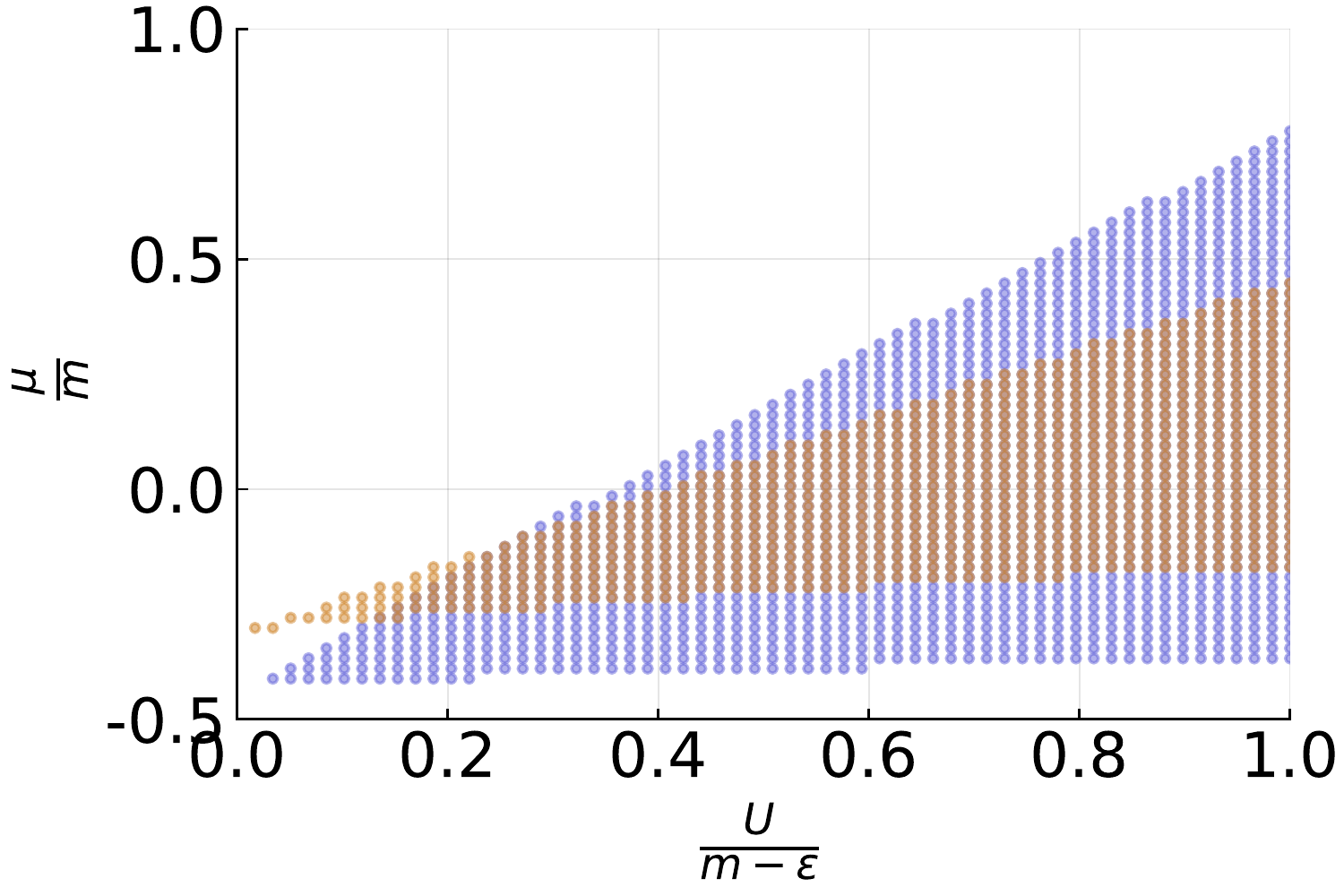}
	\\
	\includegraphics[width = 1.65in]{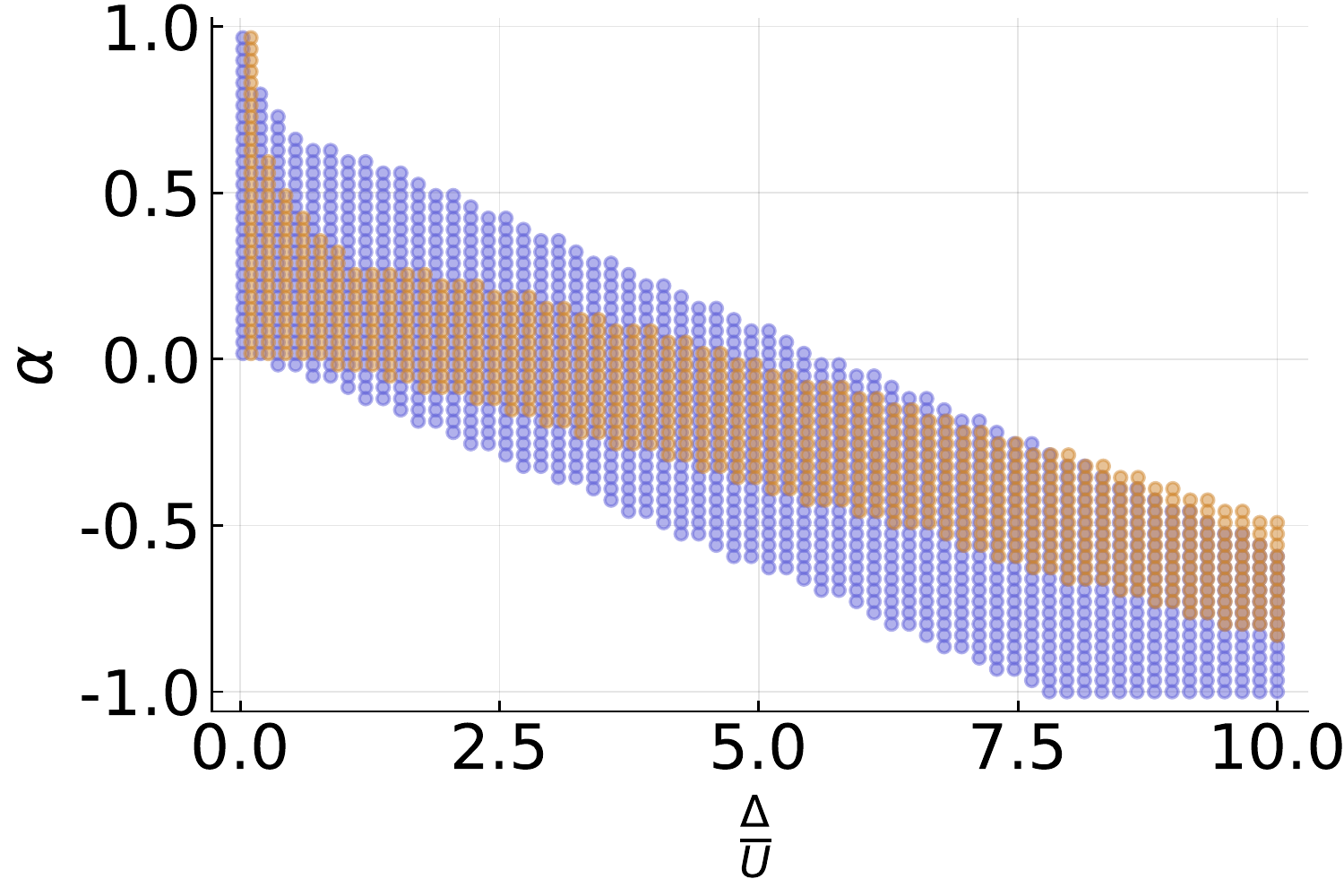}
	\includegraphics[width = 1.65in]{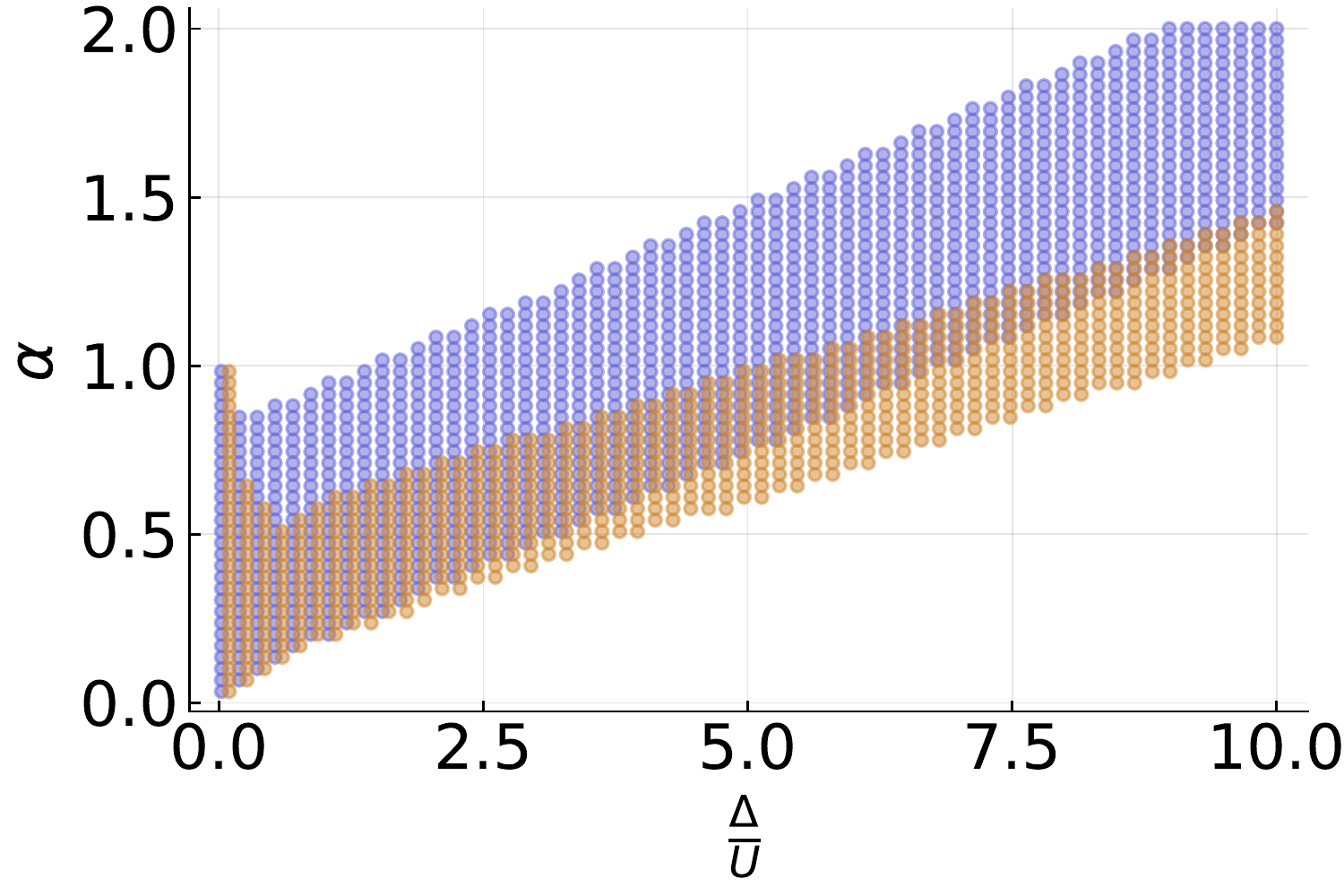}
	\caption{Phase space for magnetization for $m/D = 0.05$ and $\Delta = m/(2D)$ (blue) and $\Delta = 2m/D$ (yellow). The impurity on-site energies are $\epsilon = m/2$ (left) and $\epsilon = -m/2$ (right). For $\epsilon = m/2$, the pole shift allows the magnetization to be attained even if $\mu< \epsilon$, in agreement with Ref.~\onlinecite{Uchoa2008lms} .The finite slope of the bottom edge of the ranges in the top row is due to the spectral tails in the valence band, as explained in the text.}
	\label{fig:GapStates}
\end{figure}

While one needs to use numerical methods to obtain the self-consistent solution for the occupation numbers, it is possible to qualitatively summarize the effects that finite $\Delta$ has on the magnetic range of $\mu$ as compared to an isolated impurity. First, energy renormalization shrinks the range and moves it towards the middle of the gap. Second, the hybridization with the band states further reduces the range due to the spectral tails. Finally, because the finite $n_\text{min}$, the lower limit of the magnetic $\mu$ is $U$-dependent ($\mu > U n_\text{min}$) and for a substantially large Coulomb repulsion the ``singly"-occupied level can be pushed above the chemical potential, destroying magnetization. Since the amount of the spectral weight depends on $\Delta$, larger coupling will lead to a stronger dependence on $U$, see the top row of Fig.~\ref{fig:GapStates}.

A curious effect occurs if $\epsilon > 0$, see top left in Fig.~\ref{fig:GapStates}. Because of the pole shift towards the middle of the gap, the lower bound of magnetic $\mu$ will be below the non-shifted (isolated) $\epsilon$.

In Fig.~\ref{fig:GapStates}, we plot the magnetization phase space in two different ways: in the $\mu$-$U$ space (to focus on small values of $U$) and in $\alpha$-$\Delta / U$ space (to bring out the large-$U$ behavior), where $\alpha  = (\mu - \epsilon) / U$. As $U$ is increased past the point of $\epsilon + U > m$ in the first case, the peak of the spectral function enters the conduction band. Nevertheless, the qualitative behavior of the $\mu$ ranges does not change substantially because the separation between the charged and uncharged peaks is greater than the peak broadening.

As a final point for this configuration, note the non-monotonicity of the upper bound of the magnetic domain for $\epsilon < 0$ in the $\alpha$ - $\Delta/U$ space, also observed in Ref.~\cite{Uchoa2008lms}. This effect is a consequence of the energy renormalization: for large enough $U$, the value $\epsilon + U > 0$. With the energy renormalization, it becomes shifted towards the middle of the gap from above while $\epsilon$ moves towards the zero-energy point from below, reducing the magnetic range of $\mu$. As $U$ becomes smaller, $\epsilon + U$ ends up below the zero energy. In this case, both $\epsilon$ and $\epsilon + \mu$ get shifted in the same energy direction, leading to a straight band-like range of $\mu$.

\begin{figure}[h]
	\includegraphics[width = 1.65in]{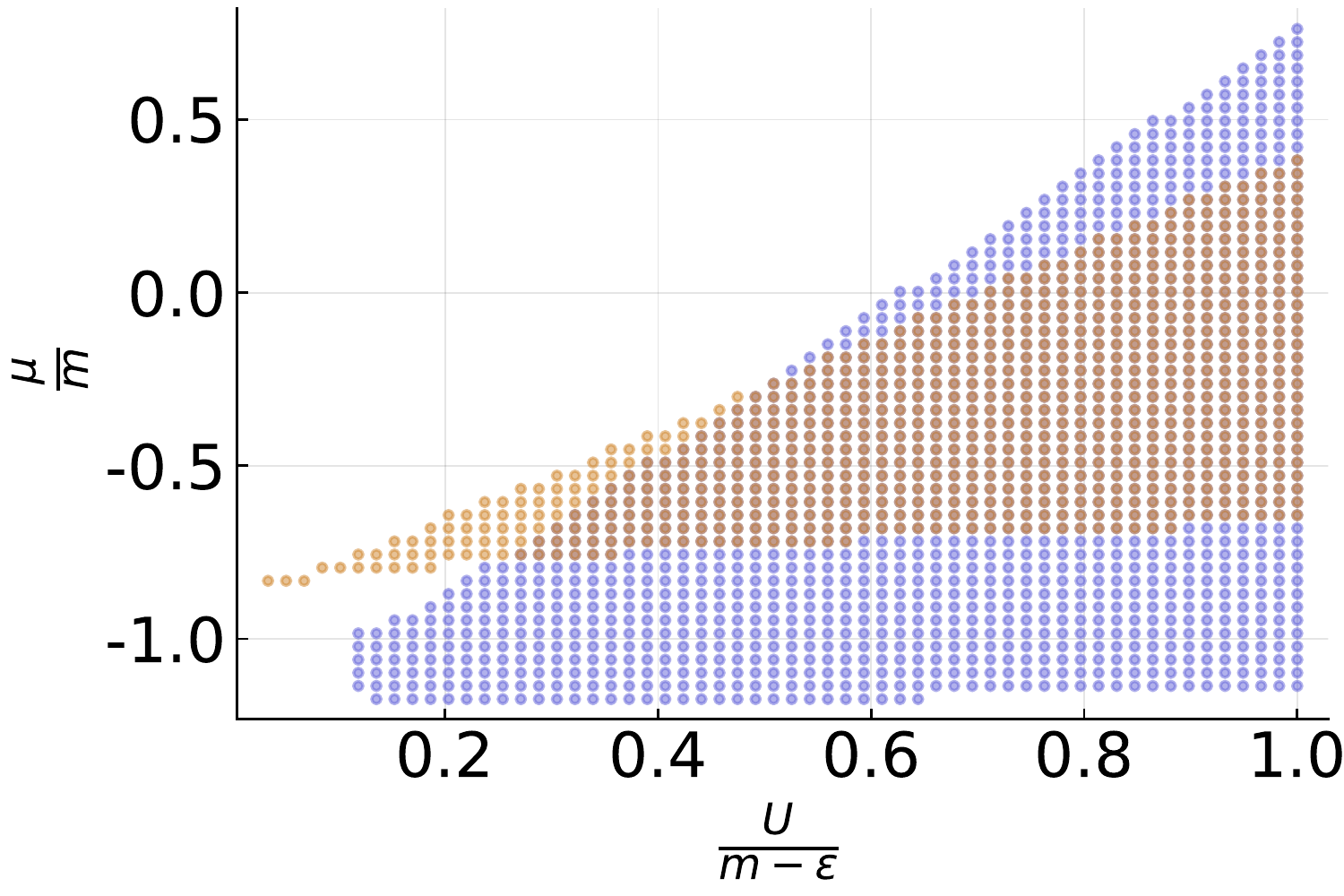}
	\includegraphics[width = 1.65in]{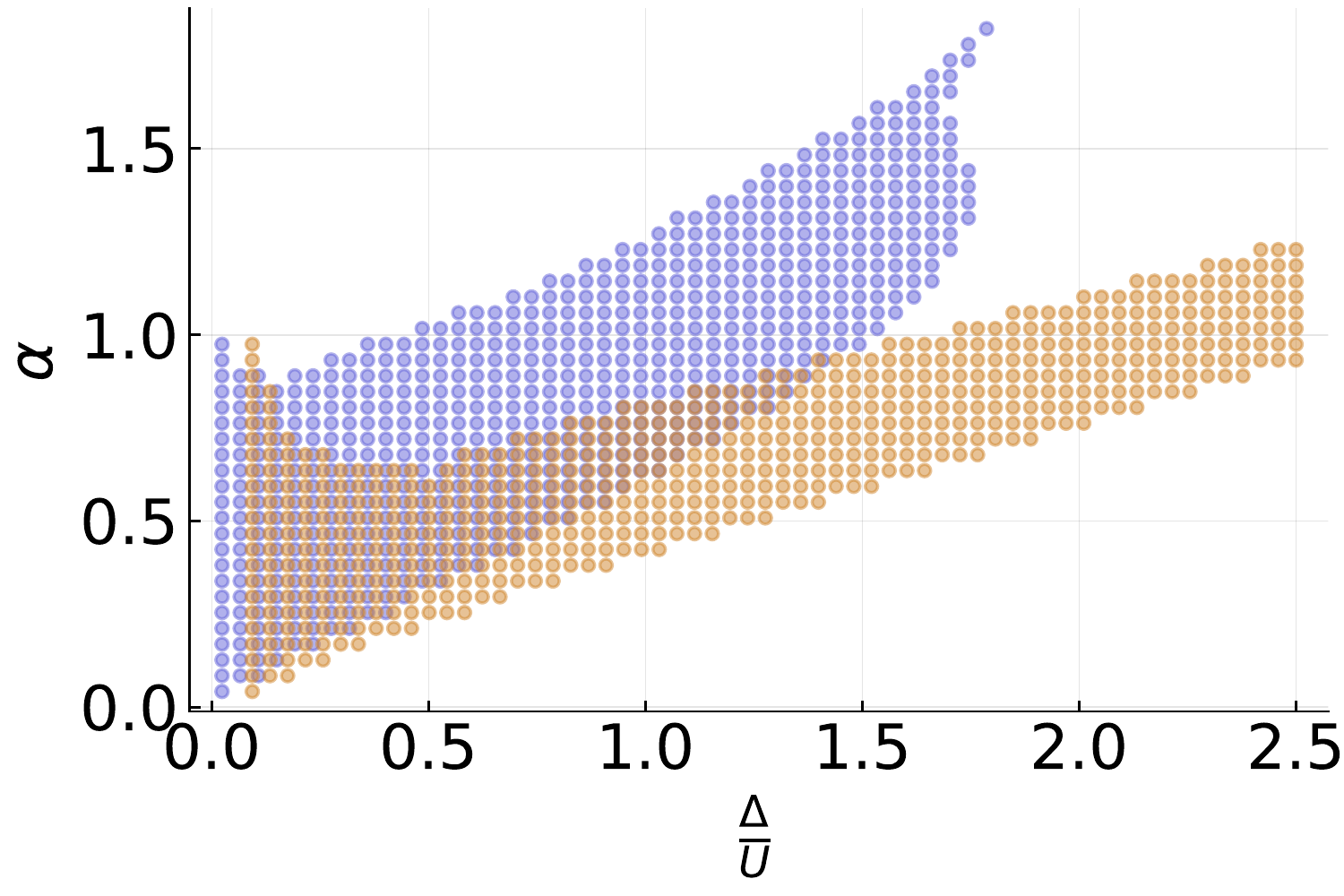}
	\caption{Magnetization for $\epsilon < -m$ with $m / D = 0.05$ and $\epsilon / m = -3/2$. The blue points and lines represent $\Delta = m/(2D)$; the yellow ones are for $\Delta = 2m/D$. The latter is sufficiently large to move the magnetic region completely into the gap. Note the similarity between the yellow plot and Fig.~\ref{fig:GapStates}.}
	\label{fig:VBStates}
\end{figure}

Next, we move to the case where $\epsilon$ is inside the energy range of the valence band. First, if the coupling parameter $\Delta$ is large enough, $\epsilon$ becomes renormalized into the gap, leaving a decaying spectral tail in the valence band. In this case, the magnetization behavior is similar to the previous section where we started with $\epsilon < 0$ inside the gap, see Fig.~\ref{fig:VBStates}. Also, if the renormalized $\epsilon$ remains in the valence band and $\epsilon + U$ is inside the gap, the qualitative behavior of the upper limit of magnetic $\mu$ is the same because it is determined by the pole crossing the chemical potential.

Even if $\Delta$ is not large enough to move the broadened peak into the gap, there is still a pole in the gap in accordance with Eq.~\eqref{eqn:P}. The pole approaches the band edge and its weight is severely reduced as $\epsilon_\sigma$ is pushed deeper into the valence band. Positioning the chemical potential between the poles for the renormalized values of $\epsilon$ and $\epsilon + U$ is similar to the $|\epsilon_\sigma|<m$ scenario which leads to a magnetizable system. Because of the very small pole weight, however, the difference between $\langle n_\sigma \rangle$ and $\langle n_{-\sigma}\rangle$ is very small, leading to a very weak magnetization. In addition, for small values of $U$, the poles are very close together, reducing the range of $\mu$ that leads to a magnetized state. The effect of this pole magnetization can be seen on the right panel of Fig.~\ref{fig:VBStates} as a ``tail" at the top of the blue plot.

When the maxima for $\epsilon$ and $\epsilon + U$ are still inside the valence band and $\mu < -m$, the poles become irrelevant and the magnetization is determined by the broadened peaks in the valence band. Because of the broadening, there is be a minimum value of $U$ required for magnetization. This effect has also been described in the earlier work~\cite{Uchoa2008lms} where $\Delta D/U$ has an upper limit for magnetization. Unlike the previous work, however, where a large coupling inevitably suppresses magnetization, here a sufficiently large $\Delta$ can relocate the entire range of magnetic $\mu$ into the gap, allowing magnetization even for infinitesimally small values of $U$.

\begin{figure}[h]
	\includegraphics[width = 1.65in]{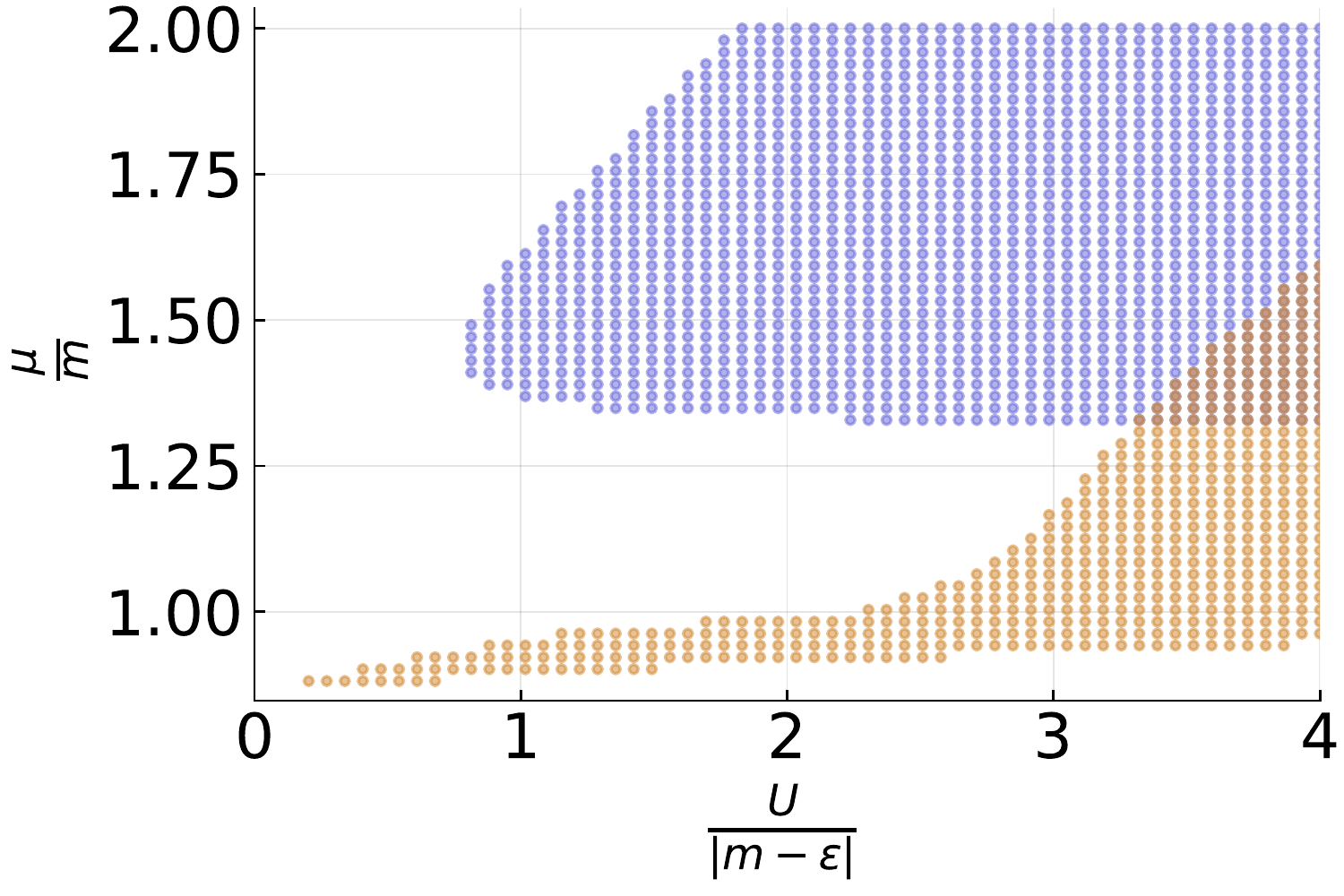}	
	\includegraphics[width = 1.65in]{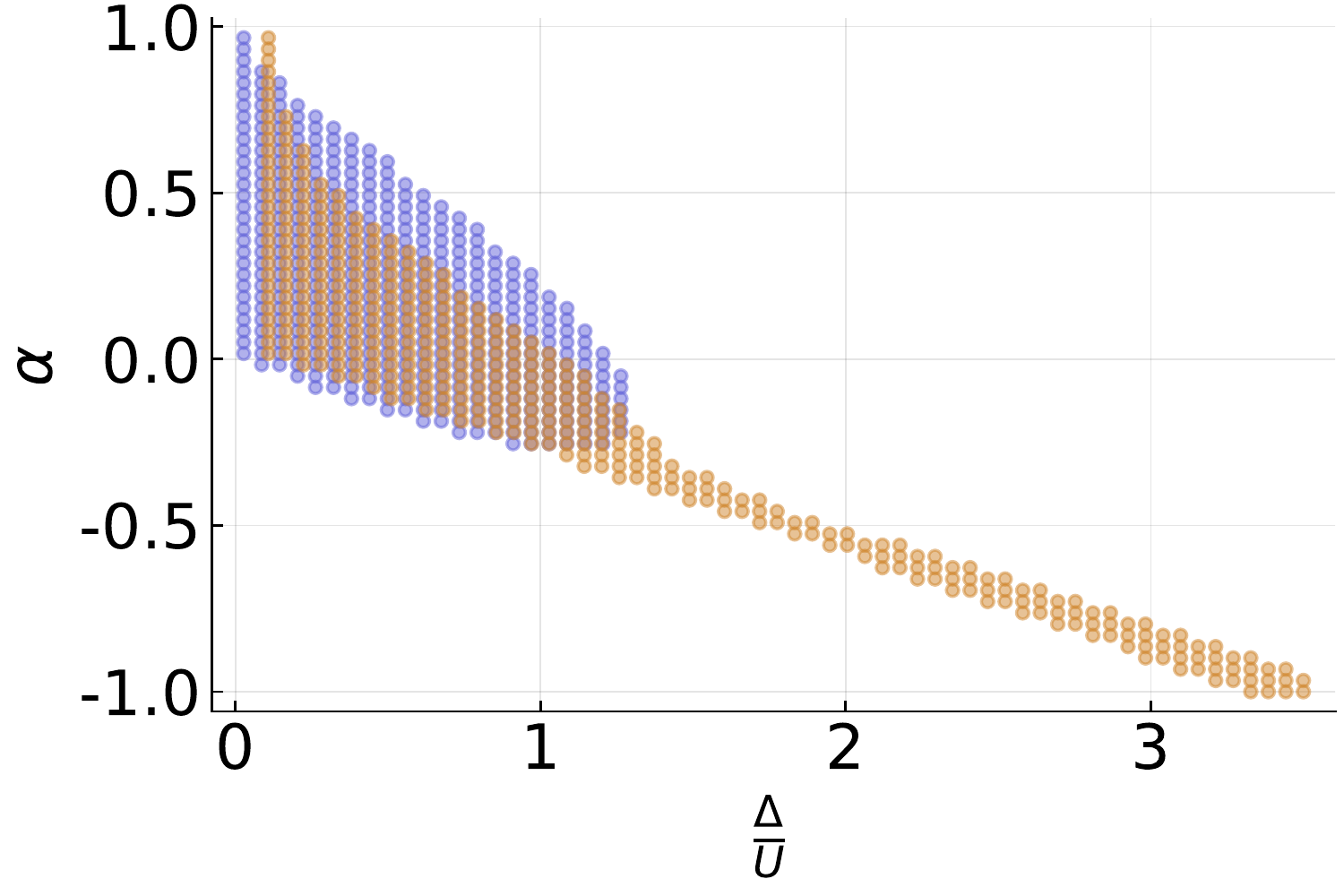}
	\caption{Magnetization for $m / D = 0.05$ and $\epsilon / m = 3/2$. The blue points represent $\Delta = m/(2D)$; the yellow ones are for $\Delta = 2m/D$.}
	\label{fig:CBStates}
\end{figure}

Finally, we turn to the case where the impurity energy lies within the conduction band, see Fig.~\ref{fig:CBStates}. If the coupling is not too strong, the range of magnetic $\mu$'s is shifted towards the middle of the gap and there is a minimum value of $U$, as expected. Because of the energy renormalization, the impurity can become magnetized even when $\mu<\epsilon$, as we saw above. For smaller value of $\Delta$, the shape of the magnetic $\mu$ region is similar to the one observed in Ref.~\onlinecite{Uchoa2008lms}.

If $\Delta$ is large enough, $\epsilon$ can be renormalized into the gap. While the renormalized value of $\epsilon + U n_\text{max}$ is also in the gap, we essentially have the case of $|\epsilon_\sigma|<m$ where the range of $\mu$ is determined by the separation between the poles and any value of $U$ can yield magnetization. As the charged energy enters the conduction band, however, the pole becomes broadened. This broadening reduces the effect that changing $\mu$ has on the occupation number and, therefore, its impact on the pole of the opposite spin in the gap. This reduced sensitivity to the chemical potential increases the magnetic range of $\mu$.

\begin{figure}[h]
	\includegraphics[width = 2.75in]{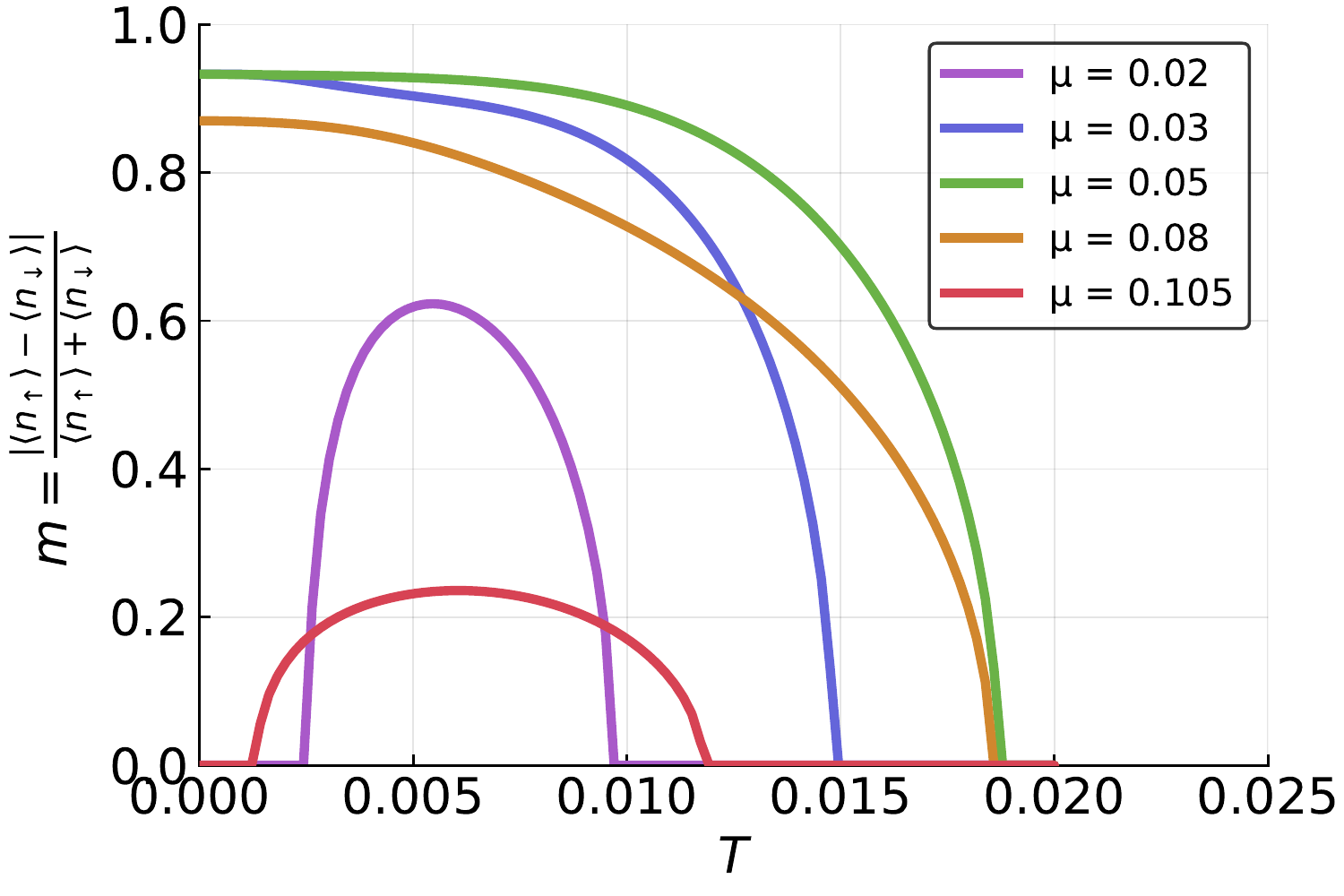}
	\includegraphics[width = 2.75in]{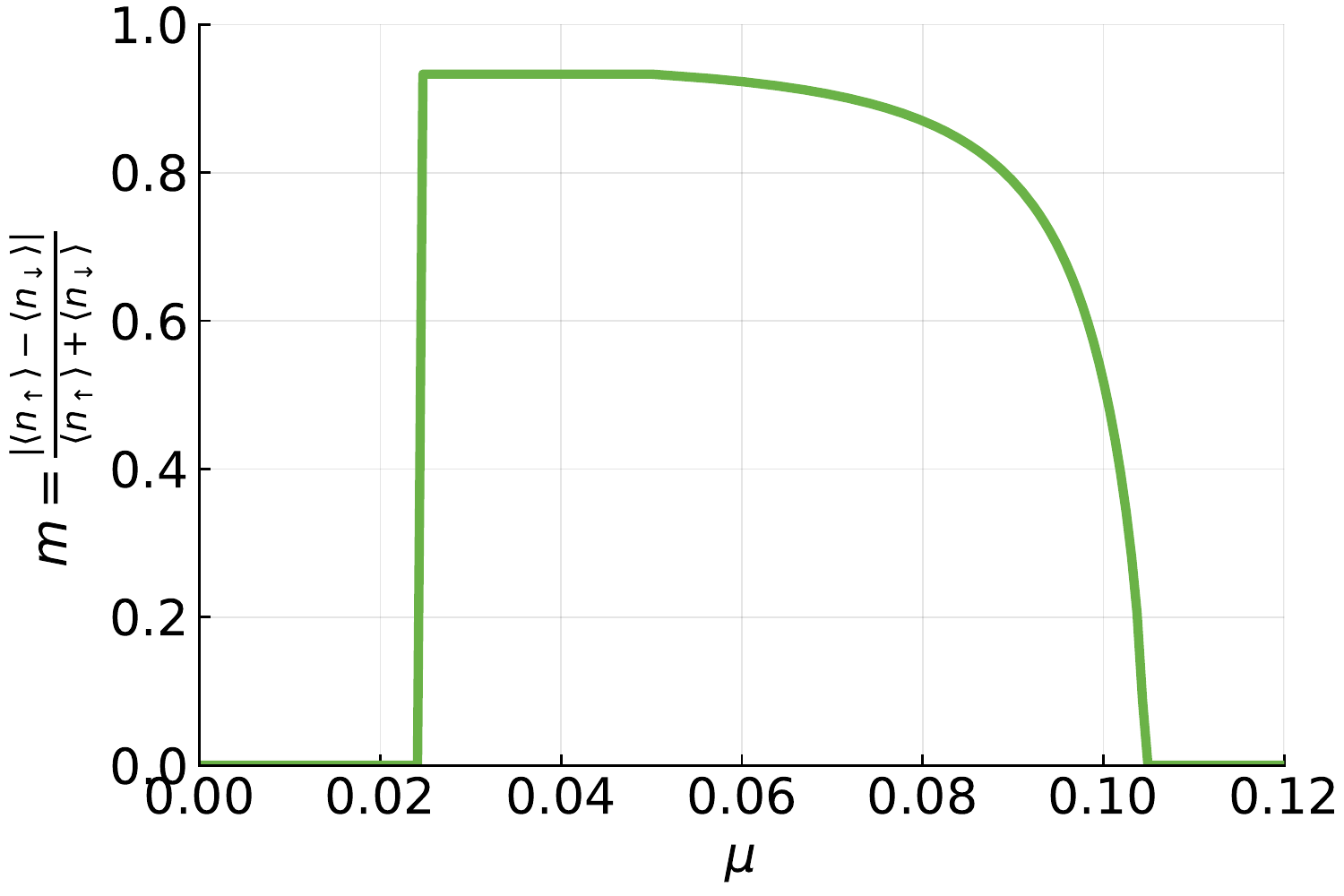}
	\caption{(Top) Magnetization as a function of temperature (given in units of $D$). (Bottom) Magnetization at $T = 0$ as a function of $\mu$.}
	\label{fig:Magnetization}
\end{figure}

So far, all the calculations were performed at $T = 0$. However, as with any magnetization problem, it is appropriate to consider the impact that finite temperature has on the system. We focus on the case where the impurity state is in the gap: $\epsilon / m = 1/2$ with $m /D = 0.05$. We choose the weaker coupling from the values we considered above and set $\Delta = m / (2D)$. Setting $U / D = 0.1$, we plot the magnetization of the system for several values of $\mu$ in Fig.~\ref{fig:Magnetization}.

For our chosen set of parameters, the pole for $\epsilon = 0.025$ is at $\approx 0.0216$. Therefore, all but one chemical potentials are above the unoccupied state. For $\mu$ below the empty state energy, the impurity is not magnetic at zero temperature, as expected. However, as the temperature is increased, a finite magnetization appears. The obvious reason is that with increasing temperature, the Fermi-Dirac distribution is no longer zero at the pole, even though it is above the chemical potential. If $U$ is large enough, the ``doubly"-occupied state is empty, leading to magnetization. Further raising the temperature destroys the magnetic state. Intermediate values of $\mu$ in Fig.~\ref{fig:Magnetization} are magnetic at zero temperature. They exhibit a standard behavior with increasing $T$, where there is a critical $T_c$ past which the system is non-magnetic.

Increasing the value of $\mu$ leads to a reentrant behavior observed at the low values of the chemical potential. However, the reason for the suppression of magnetization at low $T$ is the opposite: instead of the singly-occupied state being empty as in the low-$\mu$ case, here, at $T = 0$ even the doubly-occupied state is filled. Increasing the temperature leads to a partial vacation of the fully-filled state, allowing the system to magnetize. Unlike the low-$\mu$ behavior, this effect relies on the presence of the band edge and, therefore, cannot arise in metals.

In conclusion, we have described two tunable parameters that can be used to control magnetization in two-dimensional materials: chemical potential and temperature. We have shown that the presence of the gap not only does not preclude the appearance of magnetic states but, instead, can facilitate their formation. In fact, for certain energy configurations, even a vanishingly small on-site Coulomb repulsion can result in magnetization. This is in sharp contrast to the traditional materials where the competition between the Coulomb term and the spectral function broadening leads to suppressed magnetization. It is worth noting that while we used the massive Dirac dispersion in our work, the qualitative results are general and apply also to traditional parabolic semiconductors regardless of the dimensionality. Finally, we have described the role played by the temperature in the formation of magnetic states. In addition to traditional behavior of full magnetization at $T = 0$ and its subsequent decay with increasing temperature, we have described experimentally achievable scenarios where the magnetization exhibits a novel reentrant behavior.

The authors acknowledge the National Research Foundation, Prime Minister Office, Singapore, under its Medium Sized Centre Programme and CRP award ``Novel 2D materials with tailored properties: Beyond graphene" (R-144-000-295-281). 


%

\end{document}